\documentclass[12pt]{article}
\usepackage{amssymb,amsmath,cite,geometry,graphicx,url}
\newcommand{\diam}{\mathop{diam}}

\catcode`\@=11

\@addtoreset{equation}{section}

\long\def\@makefntext#1{\parindent 0cm\noindent
\hbox to 1em{\hss$^{\@thefnmark}$}#1}

\newcommand{\captionfonts}{\small}
\makeatletter  % Allow the use of @ in command names
\long\def\@makecaption#1#2{%
  \vskip\abovecaptionskip
  \sbox\@tempboxa{{\captionfonts #1: #2}}%
  \ifdim \wd\@tempboxa >\hsize
    {\captionfonts #1: #2\par}
  \else
    \hbox to\hsize{\hfil\box\@tempboxa\hfil}%
  \fi
  \vskip\belowcaptionskip}
\makeatother   % Cancel the effect of \makeatletter 

\begin{document}
\begin{titlepage}
\vspace{.5in}
\begin{flushright}
May 2017\\  %date
\end{flushright}
\vspace{.5in}
\begin{center}
{\Large\bf
Dimension and Dimensional Reduction\\[1.5ex]
in Quantum Gravity}\\  %title
\vspace{.4in}
{S.~C{\sc arlip}\footnote{\it email: carlip@physics.ucdavis.edu}\\
       {\small\it Department of Physics}\\
       {\small\it University of California}\\
       {\small\it Davis, CA 95616}\\{\small\it USA}}
\end{center}

\vspace{.5in}
\begin{center}
{\large\bf Abstract}
\end{center},
\begin{center}
\begin{minipage}{4.65in}
{\small
A number of very different approaches to quantum gravity contain a
common thread, a hint that spacetime at very short distances becomes effectively 
two dimensional.  I review this evidence, starting with a discussion of the physical 
meaning of ``dimension'' and concluding with some speculative ideas of what 
dimensional reduction might mean for physics.
}
\end{minipage}
\end{center}
\end{titlepage}
\addtocounter{footnote}{-1}

\section{Why Dimensional Reduction?}

What is the dimension of spacetime?  For most of physics, the answer is
straightforward and uncontroversial: we know from everyday experience
that we live in a universe with three dimensions of space and one of time.
For a condensed matter physicist, say, or an astronomer, this is simply a
given.  There are a few exceptions---surface states in condensed matter that
act two-dimensional, string theory in ten dimensions---but for the most part 
dimension is simply a fixed, and known, external parameter.

Over the past few years, though, hints have emerged from quantum gravity
suggesting that the dimension of spacetime is dynamical and scale-dependent,
and shrinks to $d\sim2$ at very small distances or high energies.  The purpose 
of this review is to summarize this evidence and to discuss some possible 
implications for physics.

\subsection{Dimensional reduction and quantum gravity}

As early as 1916,  Einstein pointed out that it would probably be necessary
to combine the newly formulated general theory of relativity with the emerging
ideas of quantum mechanics \cite{Einstein}.  In the century since,
efforts to quantize gravity have led to many breakthroughs in fundamental
physics, from gauge-fixing and ghosts to the background field method to 
Dirac's analysis of constrained Hamiltonian systems.  But the fundamental
goal of a complete, consistent quantum theory of gravity still seems distant.
In its place, we have a number of interesting but incomplete research programs: 
most famously string theory and loop quantum gravity, but also group field theory, 
causal set theory, asymptotic safety, lattice approaches such as causal dynamical
triangulations, research based on noncommutative geometry, and various ideas 
for ``emergent'' gravity.

In a situation like this, we need to explore many complementary lines of research.  
One particular strategy is to look for fundamental features that are shared by  
different quantization programs.   There is no guarantee that such features 
will persist in the ``correct'' quantum theory of gravity, but such a pattern of 
recurrence at least makes it more plausible.

We currently have one outstanding example of such a commonality, the predictions 
of black hole thermodynamics.  We have not directly observed Hawking radiation
or black hole entropy, but the thermodynamics properties of black holes can be 
derived in so many different ways, with such a variety of assumptions and 
approximations \cite{Carlipbh}, that a claim that black holes do not radiate would 
by now seem perverse.  Dimensional reduction of spacetime near the Planck scale
is a candidate for second such commonality, albeit one that is much less firmly 
established.

\subsection{Dimension as an observable}

In most of physics the dimension of space, or spacetime, is taken as a fixed 
external parameter.  While the notion of dimension is ancient---see 
\cite{Whitrow,Barrow,Calendar} for discussion of the history and 
philosophy---the mathematical formalism for a space of arbitrary dimension
is fairly recent, often attributed to Sch{\"a}fli's work in the early 1850s 
\cite{Schlafli}.  The question of \emph{why} our universe should have the 
number of dimensions it does was famously discussed in 1917 by Ehrenfest 
\cite{Ehrenfest}, who pointed out that such features as the stability of 
Newtonian orbits and the duality between electric and magnetic fields 
are unique to three spatial dimensions.  But Ehrenfest also warned that 
``the questions [of what determines the number of dimensions] have 
perhaps no sense.''

The idea that spacetime might really have more than four dimensions was 
introduced into physics by Nordstr{\"o}m \cite{Nordstrom}, and became more 
widely known with the work of Kaluza \cite{Kaluza} and Klein \cite{Klein}.
The extension beyond five dimensions first appeared, I believe, as an 
exercise in lecture notes by DeWitt \cite{DeWitt}.  Kaluza-Klein theory provides
a useful illustration of the notion that dimension might be scale-dependent.  At small 
enough distances, the spacetime of Kaluza-Klein theory is an $n$-dimensional 
manifold with $n>4$.  At larger scales, though, the compact dimensions can no
longer be resolved, and the spacetime becomes effectively four-dimensional, 
with excitations in the compact directions appearing as towers of massive 
four-dimensional modes.  The converse process of dimensional deconstruction
\cite{Arkani}, in which appropriate four-dimensional modes effectively 
``create'' extra dimensions at large distances, has more recently become 
popular in high energy theory.

Here we are interested in the opposite phenomenon, in which the number of
effective dimensions \emph{decreases} at short distances or high energies.
The first models I know of that exhibited this behavior were introduced by
Jourjine \cite{Jourjine} , Kaplunovsky and Weinstein \cite{Kaplunovsky},
Zeilinger and Svozil \cite{Zeilinger}, and Crane and Smolin \cite{Crane}, all in 
1985.   A year later, Hu and O'Connor observed scale dependence of the effective 
dimension in anisotropic cosmologies \cite{Hu}, but the wider significance was not 
fully appreciated.  In  quantum gravity, the phenomenon of dimensional reduction
first appeared in string theory, where a dimension characterizing thermodynamic 
behavior was found to unexpectedly drop to $d=2$ at high temperatures, leading 
Atick and Witten to postulate ``a lattice theory with a $(1+1)$-dimensional field 
theory on each lattice site''  \cite{Atick}. But it was only with the computation of 
the flow of spectral dimension in causal dynamical triangulations \cite{Ambjorn} 
that the idea really took hold.

To proceed further, though, we will have to first address a basic question: how, 
precisely, do we define ``dimension'' as a physical observable?

\section{Dimensional estimators}

We experience space to be three-dimensional.  We learn as children that an
object has a height, a width, and a depth (this goes back to Aristotle \cite{Whitrow}), 
and, later, that no more than three mutually perpendicular lines can be drawn from 
a point (this goes back at least to Galileo \cite{Galileo}, and perhaps to a lost work 
by Ptolemy \cite{Whitrow}).  Later, we learn that a spatial position can be specified
by three coordinates, and an event by four; that is, spacetime is homeomorphic 
to $\mathbb{R}^4$.   This picture carries over to classical general relativity, where 
spacetime is modeled by a four-dimensional manifold, that is, a structure locally 
diffeomorphic to $\mathbb{R}^4$.

But we also know this is not enough.  In mathematics, Cantor's discovery of a 
one-to-one correspondence between the points of a line segment and the points 
of a unit square showed the flaw in the intuitive idea that a two-dimensional 
space has ``more points'' than a one-dimensional space, and  Peano's construction 
of space-filling curves made it clear that dimension had to be more than a simple
counting of the number of parameters needed to specify a point \cite{Hurewicz}.   
Moreover, the physical quantity we call dimension clearly depends on scale.
We draw a line on a piece of paper and call it one dimensional, but closer up it is 
three dimensional, and even closer the ink consists of elementary particles that appear 
point-like (or perhaps string-like).   Moving to to quantum mechanics, the question 
becomes even more problematic.  Even for a single particle, the path integral is 
dominated by nowhere-smooth paths with fractal  dimensions \cite{Morette}, with
smooth one-dimensional paths appearing only semiclassically.  Quantum gravity 
adds yet another layer of complication:  in a spatially compact universe there are 
no local observables, and it is no longer obvious that a ``point'' has any real 
meaning \cite{Torre,Giddings}.  Ultimately, if our smooth spacetime is somehow 
emergent at large distances, as many approaches to  quantum gravity suggest, 
then its dimension, too, should be emergent.

How, then, do we decide what ``dimension'' means?  We need ``dimensional estimators,'' 
physical observables with a simple dependence on dimension that can be generalized 
to situations in which the meaning of dimension is ambiguous.   A number of rather
different possibilities exist, and  different choices need not always agree.  Dimension 
may depend on exactly what physical question we are asking.

Broadly speaking, dimensional estimators come in two varieties.  Some are primarily
mathematical, determining the dimension of some class of mathematical spaces
that we use to model physics.  Others are more immediately physical, starting 
with a physical quantity that has a simple dimensional dependence and using
its value to deduce dimension.  The difference is not always sharp---spectral 
dimension, for instance, is a characteristic of both a mathematical random 
walk and a physical diffusion process---but the distinction is helpful for keeping 
track of alternatives.

\subsection{Geometric dimensions \label{geomd}}

The mathematical field of ``dimension theory'' is too large to discuss in
detail in a review article such as this.  Readers may want to look at the books 
\cite{Hurewicz} or \cite{Nagami} for more formal definitions and derivations, or
at \cite{Schleicher} for an entertaining mathematical perspective.  Here I will 
restrict myself to brief summaries of the more important mathematical approaches 
to dimension, some of which have made an appearance in physics and some of 
which have not. 

\begin{itemize}
\item{\bf Topology: inductive dimension and covering dimension}\\[1ex]
Three of the common 
mathematical definitions of dimension are purely topological, depending only on
the structure of points and open sets.  The inductive dimensions originate from the 
observation that a line can be divided by a point, a surface by a curve, a three-dimensional 
space by a surface, etc.  More formally, the large inductive dimension of the empty set 
is $-1$; the large inductive dimension of a space $X$ is the smallest integer $d_I$ such 
that any disjoint pair of closed sets $F,H\subset X$can be separated\footnote{The 
sets $F$ and $H$ are separated by $K$ if $F$, $H$, and $K$ are mutually disjoint
and $X\backslash K$ consists of two disjoint sets, one containing $F$ and the
other containing $H$.}  by another closed set $K$ of large inductive dimension less 
than or equal to $d_I-1$ \cite{Nagami}.  The small inductive dimension $d_i$ is defined 
similarly, but with the closed set $H$ replaced by a single point $x$.

The covering dimension originates from Lebesgue's observation that a line can be
covered by arbitrarily small intervals in such a way that no point is contained in
more than two intervals, a square can be covered by arbitrarily small ``bricks''
in such a way that no point is contained in more than three bricks, etc.  More
formally, let $\mathcal{H}$ be an open covering of $X$, and define the order of 
$\mathcal{H}$ to be the smallest number $n$ such that each point in $X$ belongs
to at most $n$ sets in $\mathcal{H}$.  The covering dimension or Lebesgue dimension
$d_L$ is then the minimum value of $n$ such that every open cover of $X$ 
has a refinement with order less than or equal to $n+1$ \cite{Nagami}.  

By construction, the large inductive, small inductive, and covering dimensions
are integers.  For nice enough spaces---for instance, separable metric spaces---they
are all equal \cite{Hurewicz}, but there are exotic cases in which they differ.  But while 
these notions are widely used in mathematics, they have seen very little application 
to physics, except indirectly as the basis for the statement that $\mathbb{R}^n$ 
is $n$-dimensional.  The problem, I believe, is that they are too primitive and too
far removed from observation: how, for instance, do we determine physically whether 
a set of spacetime events is open?

\item{\bf Scaling: box-counting and Hausdorff dimensions}\\[1ex]
Once a notion of distance is added, mathematical approaches to dimension 
become more ``physical.''  In particular, dimension can be determined by the scaling
behavior of geometric objects.  Suppose, for instance, that a ball of radius $r$ has 
volume $V(r)$.  If at some characteristic length scale $V\sim r^n$, then $n$ is a  
measure of dimension at that scale.  This concept can be extended to graphs, 
giving upper and lower ``internal scaling dimensions'' $d_{IS}$ and $d_{is}$ 
\cite{Requardt}.  

Alternatively, suppose a ball of radius $r$ can be filled by $N(\epsilon)$ balls of 
radius $\epsilon$.  If $N(\epsilon)$ scales as $(r/\epsilon)^n$, then $n$ is again
a measure of dimension.   This idea is made precise by the notions of upper and lower 
box-counting dimension (or ``capacity'') and Hausdorff dimension.  Let $N(X,\epsilon)$ 
be the smallest number of balls of radius $\epsilon$ that can cover the set $X$.
Then if the limit exists, the box-counting dimension of $X$ is \cite{Pesin}
\begin{equation}
d_b(X) = \lim_{\epsilon\rightarrow0}\frac{\ln N(X,\epsilon)}{\ln(1/\epsilon)}
\label{a1}
\end{equation}
If the limit does not exist, one can still define upper and lower limits (the supremum
and infimum of the limit points), which determine the upper and lower box-counting
dimensions $d_C$ and $d_c$.

The Hausdorff dimension is a further refinement, in which the balls are allowed
to have different sizes.  For a metric space with a distance function $d(x,y)$, the 
``diameter'' $\diam(A)$ of a set $A$ is the largest distance between any 
two elements of $A$ (technically, the supremum $\sup\{d(x,y):x,y\in A\}$). 
Consider coverings of a set $X$ by subsets $A_i$ with $\diam(A_i)<\epsilon$.  
For $p$ an arbitrary nonnegative number, define  
\begin{align}
m(X,p) = \lim_{\epsilon\downarrow0}\inf \sum_i [\diam(A_i)]^p
\label{a2}
\end{align}
where the infimum in (\ref{a2}) is over all decompositions $X = A_1\cup A_2\cup\dots$.
As a function of $p$, $m(X,p)$ is zero for $p$ large, and jumps to infinity for
$p$ small.  The Hausdorff dimension $d_H$ is the value of $p$ at this threshold; 
that is
\begin{align}
d_H(X) = \sup\{p: m(X,p) >0\} = \inf\{p: m(X,p)<\infty\}
\label{a3}
\end{align}
The Hausdorff dimension is sometimes called the ``fractal dimension,'' as introduced 
and popularized by Mandelbrot \cite{Mandel}.

In contrast to the inductive and covering dimensions, box-counting and Hausdorff
dimensions need not be integers.   Again, though, for ``nice'' spaces---the differentiable 
manifolds of general relativity, for example---these scaling dimensions give the expected 
answer.

For a space with a measure,  generalizations of Hausdorff dimension
are available \cite{Pesin,Farmer}.  The ``information dimension,'' for example, is
a variation of the box-counting dimension adjusted to account for the relative 
probabilities of balls used to cover the set: in analogy to (\ref{a1}),
\begin{align}
d_I(X) = \lim_{\epsilon\rightarrow0}\frac{I(X,\epsilon)}{\ln(1/\epsilon)} \quad
\hbox{with}\quad I(X,\epsilon) = -\sum_{i=1}^{N(\epsilon)}P_i\ln P_i
\label{a3a}
\end{align}
where $P_i$ is the probability measure for the $i$th ball.  There are, in fact, infinitely 
many generalizations, associated with correlations of $n$ points \cite{Hentschel}.  
Yet other dimensions can be defined in terms of Lyapunov numbers of a flow 
\cite{Farmer}.  As far as I know, none of these has been used in investigations of 
quantum gravity, although the Lyapunov dimension could be relevant to the discussion 
of geodesics that will appear later in this section.

\item{\bf Random walks: spectral and walk dimensions}\\[1ex]
If the topological dimensions require too little physical information, the scaling
dimensions perhaps require too much.  It is not always easy to define a
distance function in a model of quantum gravity, so the physical meaning of a 
geodesic ball or the diameter of a set can be unclear.  An intermediate set of 
dimensions are based instead on the properties of random walks or diffusion
processes, which can be defined in more general settings.

The basic idea is straightforward.  In any space on which a random walk can 
be defined, the resulting diffusion process gradually explores larger and larger 
regions.  More dimensions mean both slower diffusion---there are more
``nearby'' points---and a slower return to the starting point.  Quantitatively, 
diffusion from an initial position $x$ to a final position $x'$ on a manifold $M$ 
is described by a heat kernel $K(x,x',s)$ satisfying
\begin{align}
\left(\frac{\partial\ }{\partial s} - \Delta_x\right)K(x,x';s) =0,
\qquad \hbox{with \quad $K(x,x';0) = \delta(x-x')$} ,
\label{a4}
\end{align}
where $\Delta_x$ is the Laplacian on $M$ at $x$, and $s$ is a measure 
of the diffusion time.  Let $\sigma(x,x')$ be Synge's world function
\cite{Synge}, one-half the square of the geodesic distance between $x$ 
and $x'$.  Then on a manifold of dimension $d_S$, the heat kernel 
generically behaves for small $s$ as \cite{DeWitt,Vassilevich,Dunne}
\begin{align}
K(x,x';s) &\sim (4\pi s)^{-d_S/2} e^{-\sigma(x,x')/2s}
    \left( 1 + [a_1]s + [a_2]s^2 + \dots \right), \nonumber\\
K(x,x;s) &\sim (4\pi s)^{-d_S/2} \left( 1 + [a_1]s + [a_2]s^2 + \dots \right)
\label{a5}
\end{align}
where the ``HaMiDeW'' coefficients $[a_i]$ are known functions of the
curvature.

Now consider any space in which a diffusion process or a random walk can be
defined.  The scale-dependent generalized spectral dimension $d_S(s)$ of a region 
$X$ is obtained from the ``return time,''
\begin{align}
d_S(s) = -2\frac{d\ln\langle K(x,x;s)\rangle_X}{d\ln s}
\label{a6}
\end{align}
where the angle brackets $\langle\ \rangle_X$ denote the average over points
$x\in X$.  For a smooth manifold, $\lim_{s\rightarrow0}d_S(s)$ gives the
ordinary geometric dimension, while for many fractals $d_S(s)$ exhibits
log periodic oscillations as $s\rightarrow0$ \cite{Dunne}.  If $s$ is small but
nonzero, $d_S(s)$ gives a scale-dependent dimension, while for 
$s$ large, $d_S(s)$ can begin to probe the global topology.

It is sometimes useful to modify the definition (\ref{a4}) of the heat kernel 
by exchanging the Laplacian $\Delta_M$ for some other physically relevant 
operator.  For some interesting cases---for example, theories of quantum gravity 
defined on manifolds but with simple nonconventional dispersion relations---%
the spectral dimension is the Hausdorff dimension of momentum space
\cite{AAGM,Calcagni}.  As a cautionary note, though, a modification of the Laplacian
can lead to a ``return probability'' that is not positive definite; in such cases,
a modification of the heat equation (\ref{a4}) may be desirable \cite{Calcagnib}.

Another dimension can also be obtained from a diffusion process.  The ``walk
dimension'' $d_W$ determines the rate at which a random walker moves away 
from the origin:
\begin{align}
d_W = 2\left(\frac{d\ln\langle x^2 K(x,0;s)\rangle}{d\ln s}\right)^{-1}  
\label{a7}
\end{align}
The walk dimension is not independent, though: it can be shown \cite{Dunne,Calcagni}
that $d_W = 2d_H/d_S$, where $d_H$ is the Hausdorff dimension.
 
\item{\bf Lorentzian dimensions:}\\[1ex]
The definitions I have discussed so far have implicitly assumed a Riemannian space,
that is, a space with a positive definite metric.  The Hausdorff dimension, for instance, 
treats spacelike and timelike distances equivalently and makes no allowance for null
separations.  The spectral dimension considers random walks without any limit
to causal paths.  There are variations, though, that take the Lorentzian nature of
spacetime into account.

Like the spectral dimension, the causal spectral dimension $d_{cs}$ \cite{Eichhorn} is based
on the behavior of random walks, but now restricted to causal paths.  Obviously, we 
can no longer use return time, since random walkers now move only forward in time.  
Instead, a dimension can be obtained from the probability that two paths starting from a 
shared initial point meet again in a diffusion time $s$.  For a manifold, the asymptotic behavior 
is obtained from the properties of biased random walks; the resulting causal spectral 
dimension is again defined by (\ref{a6}), but with $\langle K(x,x;s)\rangle$ replaced by 
the meeting probability $P_{\hbox{\tiny meet}}(s)$.

Two other Lorentzian dimensions, both based on scaling, have their origin in causal set
theory \cite{Reid}.  The first, the Myrheim-Meyer dimension \cite{Myrheim,Meyer}, 
is a kind of box-counting dimension in which the ``boxes'' are Alexandrov neighborhoods, 
or causal diamonds.  We start with $d$-dimensional Minkowski space, and pick two
causally related points $p$ and $q$ with $p\prec q$ (where $\prec$ means ``is in
the past of'').  The Alexandrov interval $I[p,q]$ is the intersection of the future of $p$
and the past of $q$.  A point $r$ in $I[p,q]$ (that is, for which $p\prec r \prec q$)
forms two new smaller intervals, $I[p,r]$ and $I[r,q]$.  It is not hard to show that
these intervals have volumes that obey a scaling relation  
\begin{align}
\frac{\langle\mathrm{Vol}(I[r,q])\rangle_r}{\mathrm{Vol}(I[p,q])}
   = \frac{\Gamma(d+1)\Gamma(\frac{d}{2})}{4\Gamma(\frac{3d}{2})}
\label{a8}
\end{align}
where the  bracket denotes an average over $r$.
The right-hand side of (\ref{a8}) is monotonically decreasing with $d$, so the
formula can be inverted to determine a dimension.  Generalizing to arbitrary
spacetimes in which causal relations are defined---including even discrete 
spacetimes such as causal sets---we obtain a Lorentzian scaling dimension, the
Myrheim-Meyer dimension $d_{MM}$.  Further curvature corrections can be added 
if the manifold is not flat \cite{Sinha}.

The second dimensional estimator coming from causal set theory is the midpoint scaling
dimension \cite{Sorkin}.  Again consider an Alexandrov interval $I(p,q)$, and define the 
``midpoint'' to be the point $r$ for which $I(p,r)$ and $I(r,q)$ are most nearly equal, that is,
the point that maximizes $N' = \min\{\mathrm{Vol}(I(p,r)),\mathrm{Vol}(I(r,q))\}$.
Then $d_{mid} = \log_2[\mathrm{Vol}(I(p,q))/N']$ gives an estimate of the dimension.

\item{\bf Geodesics:}\\[1ex]
A final, more qualitative estimate of dimension can be obtained by examining the behavior
of geodesics on a manifold.  This idea goes back to Galileo's observation that in a
three-dimensional space, at most three perpendicular lines can be drawn from a point 
\cite{Galileo}.  In a $d$-dimensional Lorentzian manifold, $d-1$ orthogonal timelike 
geodesics can leave a point.  But these geodesics may not behave identically; they 
may probe very different distances in different directions.  

For instance, let us choose Gaussian normal coordinates, select two random nearby 
geodesics at $t=0$, and look at their proper distance in each of $d-1$ orthogonal directions 
at a later time $t$.  In regions of certain manifolds---for instance, in Kasner space for both 
small and large $t$ \cite{Carlipa,Carlipb}---a random geodesic will ``see'' fewer than the 
full $d-1$ dimensions; that is, the proper distance will be large in $d_{geod}-1$ dimensions 
and much smaller in the rest.  If we consider a measurement with a finite resolution, 
the manifold will thus appear to have $d_{geod}$ dimensions.  This behavior was first 
observed by Hu and O'Connor in cosmology \cite{Hu}.  They were considering large scale
behavior, and called $d_{geod}$ the ``infrared dimension,'' but as we shall see later, a 
similar phenomenon may happen at small scales.

Since the heat kernel is determined by random walks, one might expect this phenomenon to
also be apparent in the spectral dimension.  In a sense, it is.  At a given spacetime point, 
the small $s$ expansion (\ref{a5}) is dominated by the first term, which gives the ordinary
spectral dimension.  But for a fixed $s$---a fixed ``resolution'' of the diffusion process---it 
may happen that higher $[a_i]$ terms dominate in certain regions of spacetime, lowering
the spectral dimension (\ref{a6}).  The Lyapunov dimension of the geodesic flow, mentioned 
briefly above, might be helpful in understanding this behavior more quantitatively.
 
\end{itemize}

\subsection{Physical dimensions}

The dimensional estimators of the preceding section were based on characteristics of
particular mathematical models of spacetime.  We now turn to estimators based on
the behavior of matter in spacetime.  The distinction is not sharp, of course---the
spectral dimension, for instance, can characterize either an abstract random
walk or a physical diffusion process---but it can be helpful in clarifying the meanings
of certain dimensional estimators.

\subsubsection{Thermodynamic dimensions \label{thermd}}

Many physical dimensional estimators are based on thermodynamic properties.  The
starting point is the simple observation that the density of states, and therefore
the partition function, depends on the phase space volume, and that this, in turn,
depends on dimension.  In particular, for a free particle at high energy in $d$ spacetime 
dimensions,
\begin{align}
\rho(E)dE \sim \frac{V_{d-1}}{\hbar^{2(d-1)}}E^{d-2}dE
\label{b1}
\end{align}
from which the partition function $Z(\beta)$ for a collection of noninteracting particles
and all of the associated thermodynamic quantities can be determined.

\begin{itemize}
\item{\bf Free energy:}\\[1ex]
For a free field or an ideal gas in a box of volume $V_{d-1}$ in a flat $d$-dimensional 
spacetime, the free energy is  
\begin{align}
F(\beta) = - \frac{1}{\beta}\ln Z(\beta) 
   = \pm \hbox{\it const.}\,V_{d-1}\int\frac{d^{d-1}k}{(2\pi)^{d-1}}\frac{1}{\beta}%
   \ln\left( 1 \mp e^{-\beta\omega(k)}\right)  \underset{\beta\rightarrow0}{\sim} V_{d-1}T^d
\label{b2}
\end{align}
where $\beta$ is the inverse temperature and the top sign is for bosons, the bottom for fermions.   
The temperature dependence provides a simple thermodynamic dimensional estimator $d_{th1}$.  
It is easy to see that this quantity is sensitive to  the physics: if one changes the dispersion 
relations---that is, the relationship of $\omega(k)$ to $k$---then in general $d_{th1}$ will 
change as well \cite{Nozari}.
\item{\bf Internal energy and equation of state}\\[1ex]
A number of thermodynamic quantities can be derived from the free energy.  In particular,
the energy density at high temperature is
\begin{align}
\rho = -\frac{T^2}{V_{d-1}}\frac{\partial\ }{\partial T}\left(\frac{F}{T}\right) 
    \underset{\beta\rightarrow0}{\sim}  (d-1)T^d
\label{b3}
\end{align}
This is essentially the Stefan-Boltzmann law, and it defines a new estimator $d_{th2}$.  The
pressure at high temperature is
\begin{align}
p = T\frac{\partial\ }{\partial V_{d-1}}\left(\frac{F}{T}\right)  \underset{\beta\rightarrow0}{\sim} T^d
\label{b4}
\end{align}
and the ratio of pressure to energy density determines the equation of state parameter
\begin{align}
w  \underset{\beta\rightarrow0}{\sim} 1/(d-1)
\label{b5}
\end{align}
giving another estimator $d_{th3}$.   For more detailed models of dimensional reduction, 
one can sometimes carry this approach further and compute the effect of fractional dimensions
on the full black body spectrum \cite{Caruso}.  Although the thermodynamic dimensional 
estimators agree for ordinary thermodynamics in flat spacetime, there are known examples 
in which they different from each other \cite{Amelino,Husain}.   
\item{\bf Equipartition}\\[1ex]
One more thermodynamic dimensional estimate comes from the equipartition theorem,
which tells us that a system at thermal equilibrium has an energy of $\frac{1}{2}T$ per
degree of freedom.  In particular, an ideal monoatomic gas in $d$ spacetime dimensions
has $d-1$ translational degrees of freedom per atom, so
\begin{align}
E = \frac{d-1}{2}NT ,\qquad C_V = \frac{d-1}{2}
\label{b6}
\end{align}
where $C_V$ is the heat capacity at constant volume.  This gives a final thermodynamic
dimensional estimator, $d_{th4}$, whose meaning is perhaps the most transparent:  
as a direct count of translational degrees of freedom, it is the thermal analogy of
Galileo's counting of perpendicular lines from a point.
\end{itemize}

\subsubsection{Other physical dimensions \label{otherd}}

While thermodynamic quantities provide convenient estimators of dimension, they
are by no means the only physical quantities with simple dimensional dependence.
At least three others have been considered.
\begin{itemize}
\item{\bf Greens functions}\\[1ex]
As Ehrenfest stressed in his seminal paper on the dimension of space \cite{Ehrenfest}, 
the Newtonian gravitational potential in a $d$-dimensional spacetime varies as
$r^{-(d-3)}$.  A slightly more invariant version of this statement is that the short distance 
Hadamard Greens function for massless particles has the form 
\begin{align}
G^{(1)}(x,x') \sim \left\{ \begin{array}{lc} \sigma(x,x')^{-(d-2)/2} \quad& d>2\\
                              \ln\sigma(x,x') & d=2 \end{array}\right.
\label{c1}
\end{align}
where the world function $\sigma(x,x')$ is half the squared geodesic distance 
between $x$ and $x'$ \cite{Synge}.  The generalization to an arbitrary spacetime
provides a new dimensional estimator $d_G$.  This quantity is closely related to the
spectral dimension, since the Greens function is a Laplace transform of the heat
kernel, but there can be subtle differences that depend on exactly which Laplacian
is used.  In causal set theory, for instance, the causal spectral dimension $d_{cs}$ 
has a behavior quite differently from that of the dimension $d_G$ determined by the 
Laplacian with the proper classical limit \cite{Eichhorn,Carlipc,Belenchia}.  

The use of tis estimator forces us to address a puzzling question about dimensional 
reduction and unitarity.  In ordinary quantum field theory, positivity of the spectral 
function $\rho(\mu)$ in the K{\"a}llen-Lehman representation of the two-point function
implies that the propagator in a four-dimensional spacetime cannot vanish faster
than $1/p^2$ at large $p$ \cite{Weinberg}.  Thus in field theoretic approaches
such as asymptotic safety, and in discrete approaches that have something like a
K{\"a}llen-Lehman representation, one must worry that dimensional reduction
could indicate  the appearance of  negative-norm states or a breakdown in 
quantum field theory \cite{Belenchia,Becker}.   There are several ways to evade 
this problem, typically involving states that appear in the spectral decomposition 
but are not present asymptotically, but any theory of quantum gravity that predicts
dimensional reduction will eventually have to address this issue.

\item{\bf Unruh dimension}\\[1ex]
As Unruh showed long ago \cite{Unruh}, an accelerated detector will detect particles
even in the Minkowski vacuum.  The detection rate is determined by the two-point
function of the particle being detected, and as we have just seen, such a two-point function 
has a characteristic dependence on dimension.  Unruh radiation thus gives a new,
slightly different probe of the behavior of Greens functions.  Alkofer et al.\ propose 
in \cite{Alkoferb} to use the response function of such a detector---in principle a 
directly measurably quantity---to define an ``Unruh dimension'' $d_U$.

\item{\bf Scaling dimensions and anomalous dimensions}\\[1ex]
Physical fields have natural scaling dimensions, which describe their changes under
constant rescaling of lengths and masses.  A scalar field $\varphi$, for instance,
has a term in its action of the form $(\partial\varphi)^2$, integrated over a
$d$-dimensional spacetime.  For the action to be invariant under constant changes
in scale  (in units $\hbar=1$),\footnote{In quantum field theory it is conventional to 
describe fields in terms of mass dimension rather than length dimension, hence the 
$-\Delta$ rather than $\Delta$}  $\varphi$ must scale as $L^{-\Delta_0}$ with 
$\Delta_0 = \frac{d-2}{2}$.  The behavior (\ref{c1}) of the two-point function of a 
scalar field can be read off from this scaling.
 
In an interacting quantum field theory, however, the ``canonical dimension'' 
(or ``engineering dimension'') $\Delta_0$ typically receives corrections  from 
renormalization.  The full scaling dimension becomes $\Delta = \Delta_0 + \gamma(g)$, 
where $\{g\}$ are the coupling constants and $\gamma(g)$ is called the anomalous 
dimension \cite{RG}.  The anomalous dimension changes with energy scale under
 the renormalization group flow, so $\Delta$ naturally defines a scale-dependent 
 dimension.  Note, though, that the anomalous dimension of a composite operator 
 $\mathcal{O}_1\mathcal{O}_2$ is not, in general, just a sum $\Delta(\mathcal{O}_1)
 +\Delta(\mathcal{O}_2)$, so different choices of operators may give different dimensional 
 estimators.  One particularly natural choice is to look at the dimension of a two-point 
 function $\langle0|\varphi(x)\varphi(x')|0\rangle$ of a scalar field; this gives a quantum
version $d_{qG}$ of the Greens function dimension $d_G$ considered above.

\end{itemize}

\section{Evidence for dimensional reduction \label{ev}}

We can now turn to our main topic, the hints of short distance dimensional
reduction in quantum gravity.  This is a bit tricky, since we do not yet have a
complete theory of quantum gravity; nor, as we have seen, do we have a unique
way to define dimension.  Still, by looking at a variety of approaches to quantum
gravity and dimension, let us see how far we can get.

\subsection{High temperature string theory \label{str}}

The first indication of dimensional reduction in quantum gravity came from the study
of high temperature string theory \cite{Atick}.  As a gas of strings is heated, it
undergoes a phase transition, the Hagedorn transition.  In 1988, Atick and Witten 
found that above the Hagedorn temperature, the number of degrees of freedom
abruptly drops: the free energy goes as $F/V\sim T^2$, so the thermodynamic
temperature $d_{th1}$ of section \ref{thermd} falls to $d_{th1} = 2$.   Other
thermodynamic quantities are not directly computed in \cite{Atick}, but it is easy to 
check that the dimensions $d_{th2}$ and $d_{th3}$ also fall to $d=2$.  Atick and 
Witten conclude that ``the mysterious system that prevails under distances
 $\sqrt{\alpha'}$'' behaves ``as if this system were a (1+1)-dimensional field theory.'' 

This dimensional reduction occurs for strings in a flat background.  It may  
differ in other settings: for instance, the radial dependence of temperature
in an AdS black hole background can alter the structure of the transition in
a way that changes $d_{th1}$ \cite{Andreev}.  So it is interesting to see whether
there are other indications of short distance dimensional reduction in string theory.  
This is not so easy.  As Gross and Mende stress \cite{Gross}, we cannot  consistently 
introduce external probes into string theory to explore short distance behavior.  If,
on the other hand, we use a string as a probe, we find that it sees a region 
with a size of order $(\ln E)^{1/2}$ at high energies, limiting the scale that
can be explored \cite{Susskind}.  

It is known that the amplitude for two-string scattering at fixed small angle falls 
off exponentially with energy \cite{Gross,Ooguri}, much faster than in any ordinary 
quantum field theory.  This \emph{might} be taken as a hint of dimensional reduction; 
Gross has speculated that this behavior and related high energy symmetries could 
indicate that the only nontrivial scattering in the high-energy limit is (1+1)-dimensional 
\cite{Grossb}.  As I will discuss in section \ref{silence}, this behavior may be connected to 
the phenomenon of ``asymptotic silence'' that appears in other approaches to 
dimensional reduction.  It also seems that at leading order in high energy open string 
scattering, only polarizations in the plane of the scattering are important \cite{Manes}; as
Gross and Ma{\~n}es put it, this ``suggests that the strings are trying to effectively 
reduce the number of spatial dimensions that they live in at high energy.''  Still, though, 
for now these are only hints of interesting behavior.

\subsection{Causal dynamical triangulations \label{CDT}}

The results from string theory were intriguing, but dimensional reduction 
only became a significant focus of research in 2005, with the discovery of  
short distance reduction of the spectral dimension in causal dynamical triangulations 
\cite{Ambjorn}.  Causal dynamical triangulations is a discrete approach to quantum 
gravity in which curved spacetimes are approximated by piecewise flat simplicial
complexes, whose contributions are combined, typically numerically, to form 
a path integral.  The action for a simplicial complex, the Regge action, has
been known since 1961 \cite{Regge}, and the idea of combining Regge calculus 
with Monte Carlo methods to evaluate the path integral on a computer 
dates back to  1981 \cite{Rocek}.  Until rather recently, though, no well-behaved 
continuum limit had been found; instead, the simulations typically yielded a 
``crumpled'' phase with very high Hausdorff dimension and a 
two-dimensional ``branched polymer'' phase \cite{Loll}.  The causal dynamical
triangulations program added a new ingredient, a fixed ``direction of time,''%
\footnote{The direction of time is usually introduced by fixing a time-slicing 
and demanding that all simplices connect one slice to the next, but there is 
evidence that such a strong slicing condition is not needed as long as a fixed
causal structure is maintained \cite{Lollb}.} which controls fluctuations of
topology and suppresses the unphysical phases.   The program has been
remarkably successful, producing not only a de Sitter ground state with the
correct volume profile but also the expected spectrum of quantum volume fluctuations 
\cite{Ambjornb} and reasonable transition amplitudes \cite{Cooperman}. 

The construction of a four-dimensional spacetime in causal dynamical triangulations
starts with four-dimensional simplices.  But this does not necessarily determine
the large scale spacetime dimension; the path integral may be dominated by configurations 
that are not manifold-like or that have very different macroscopic appearances, and 
ultimately the dimension has to be measured.  It is easy to define a random walk on 
a simplicial complex, so a natural choice for a dimensional estimator is the spectral 
dimension $d_S$ of section \ref{geomd}.  In 2005, Ambj{\o}rn, Jurkiewicz, and Loll 
found, quite unexpectedly, that while the expectation value of the spectral dimension 
is $d_S=4$ at large scales---the model reproduces a four-dimensional  universe at 
large distances---it drops to $d_S\approx2$ at small scales \cite{Ambjorn}.  Similarly, 
if one starts with a three-dimensional model the spectral dimension drops from 
$d_S=3$ at large scales to $d_S\approx2$ at small scales \cite{Benedetti}.  These 
results have been checked independently \cite{Kommu,Coumbe}, and verified 
analytically in a simplified toy model \cite{Giasmidis}.  Because of numerical 
uncertainty, the lower limit is not known exactly, and may be consistent with 
$d_S=1.5$ \cite{Coumbe}, but the occurrence of short distance dimensional 
reduction in this approach is unambiguous.

Very recently, there has also been renewed interest in the ``old-fashioned'' Euclidean 
dynamical triangulations approach, in which no causality condition is imposed.  While 
this usually yields phases that look nothing like our spacetime, there are indications that 
a physically reasonable continuum limit may exist if one fine-tunes the path integral 
measure \cite{Coumbex}.  The resulting spacetimes have spectral dimensions that 
falls from $d_S\approx4$ at large scales to $d_S\approx1.5$ at small scales, although 
the simulations so far are quite coarse, leading to large uncertainties in these numbers.

\subsection{Asymptotic safety \label{asafe}}

Viewed as an ordinary quantum field theory, general relativity is nonrenormalizable:
its effective action has an infinite number of terms, containing arbitrarily high powers
of curvature, each with its own coupling constant.  Even if all but a few of the
coupling constants are set to zero at some scale, they will reappear at other scales
under the renormalization group flow.  As Weinberg first pointed out, though, the
theory might still make sense \cite{Weinbergb}.  For the theory to remain valid at
arbitrarily high energies or short distances, the renormalization group flow must 
have an ultraviolet fixed point.  If it does, and if that fixed point has only a finite
number of relevant directions, then the infinitely many couplings of the low energy 
theory would be determined by finitely many high energy parameters.  Such a 
theory is ``asymptotically safe.''

Whether general relativity is asymptotically safe is a deep question, and we are 
far from a definitive answer.  But several pieces of evidence, coming from 
truncations of renormalization group equations and from exact calculations in 
dimensionally reduced models, make it plausible that it is \cite{Reuter,Nieder,Falls}.  
If gravity can be described by an asymptotically safe theory with a non-Gaussian 
ultraviolet fixed point, the anomalous dimensions introduced in 
section \ref{otherd} will flow to fixed values under the renormalization group.  For us, 
the key point is that operators at the fixed point will \emph{necessarily} acquire 
precisely the anomalous dimensions necessary to make the theory appear two-dimensional 
\cite{Lauscher,Percacci,Litim}.  In particular, the dimension $d_{qG}$ of section 
\ref{otherd} will flow to $d_{qG}=2$ at the fixed point.  

Intuitively, this happens because the theory becomes scale invariant at a fixed point, 
and the Einstein-Hilbert action is scale invariant only in two dimensions.  The more formal 
demonstration of this behavior is fairly straightforward \cite{Nieder,Litim}, although there 
are a few subtleties \cite{Percacci}.  Consider the renormalization group flow of the 
dimensionless coupling constant $g_N(\mu) = G_N\mu^{d-2}$, where $G_N$ is Newton's 
constant and $\mu$ is the mass scale.  Under this flow, we have
\begin{align}
\mu\frac{\partial g_N}{\partial\mu} 
   = [d-2+\eta_N(g_N,\dots)] g_N  ,
\label{d1}
\end{align}
where the anomalous dimension $\eta_N$ depends on both $g_N$ and any other 
(dimensionless) coupling constants in the theory.  It is evident that a free field,
or ``Gaussian,'' fixed point can occur at $g_N=0$.  For an additional non-Gaussian 
fixed point $g_N^*$  to be present, though, the right-hand side of (\ref{d1}) must vanish: 
$\eta_N(g_N^*,\dots) = 2-d$. 

But the momentum space propagator for a field with an anomalous dimension $\eta_N$ 
goes as $(p^2)^{-1 + \eta_N/2}$.  For $\eta_N = 2-d$, this becomes $p^{-d}$, 
and the associated position space propagator depends logarithmically on distance.  As
we saw in section \ref{otherd}, such a  logarithmic dependence is indicative of a 
two-dimensional conformal field, $d_{qG}=2$.  A variation of this argument shows 
that matter fields interacting with gravity at a non-Gaussian fixed point exhibit 
two-dimensional behavior as well \cite{Nieder}.  
 
We can also apply renormalization group methods to the spectral dimension $d_S$ near
a fixed point \cite{Reuterb,Rechenberger}.  The renormalization group flow effectively 
gives a scale dependence to the metric in the heat equation, inducing a flow of the return 
probability $K(x,x;s)$  of eqn.\ (\ref{a6}).  Using two different truncations of the effective 
action, it has been confirmed that three scaling regimes appear: a classical regime with 
$d_S = 4$, a quantum regime near the fixed point with $d_S=2$, and an intermediate
semiclassical regime with a spectral dimension that depends on the truncation and the
renormalization group trajectory.  Efforts are now underway to evaluate the anomalous 
dimensions of length and volume operators \cite{Reuterc}, which might eventually lead 
to results for the scaling dimensions of section \ref{geomd} near the fixed point.

\subsection{Causal set theory \label{CST}}

Yet another sign of dimensional reduction comes from a completely different
approach to quantum gravity.  Causal set theory \cite{Bombelli}  starts with perhaps
the most primitive picture of a spacetime, a collection of discrete points whose
only relations are those of causality.  Most such sets are not at all manifold-like,
but the hope is that a suitable dynamical principle will pick out the physically
relevant class. For those causal sets that \emph{do} approximate manifolds, the 
Lorentzian metric structure is determined by the causal structure, which fixes
the conformal class of the metric \cite{Malamet}, and the number of points in a 
region, which determines the conformal factor.

Causal sets are intrinsically Lorentzian, and we should use dimensional estimators
that respect this property.   One natural choice is the Myrheim-Meyer dimension 
$d_{MM}$ introduced in section \ref{geomd}.  It has been shown that for very 
small causal sets---those with four, five, or six elements---the average Myrheim-Meyer 
dimension is $d_{MM}\approx 2$ \cite{Carlipc}.  Work on extending this computation
to larger sets is currently in progress.  Further evidence comes from ``generic'' causal 
sets, which have a structure known as a Kleitman-Rothschild (KR) order \cite{KR}.  Such sets 
are not at all manifold-like, and must be dynamically suppressed at large scales, 
but they might arguably remain important at small distances.  KR orders have a Myrheim-Meyer 
dimension of $d_{MM} = 2.38$.  There are also indications that causal sets obtained 
by randomly ``sprinkling'' points in Minkowski space have Myrheim-Meyer dimensions 
that fall to $d_{MM}\approx 2$ for small subspaces \cite{Reid}; again, work on this 
question is in progress.

We can also consider the causal spectral dimension $d_{cs}$ of section \ref{geomd}.
Here, rather dramatically, Eichhorn and Mizera have shown that the dimension of a 
manifold-like causal set \emph{increases} at short distances \cite{Eichhorn}.  However, 
the d'Alembertian implicit in the definition of $d_{cs}$ also fails to reproduce the standard 
flat spacetime d'Alembertian on causal sets that approximate Minkowski space.  If one 
instead computes the spectral dimension from the ``right'' d'Alembertian, one obtains 
a result that shows the usual pattern of dimensional reduction to $d_{cs}=2$ at short
distances \cite{Belenchia}.  It also seems that the Greens function dimension $d_G$ of 
section \ref{otherd} falls to $d_G=2$ at short distances, although there are some 
ambiguities in a choice of regularization \cite{Carlipc,Belenchia}.

\subsection{Loop quantum gravity}

Another place that we might look for evidence for dimensional reduction is loop
quantum gravity, a popular canonical quantization scheme based on a particular choice of
connection variables and a consequent nonstandard inner product \cite{Rovelli}.  At the 
kinematical level, loop quantum gravity has been very successful, with a rigorously defined 
Hilbert space of spin network states and a collection of interesting geometrical operators.  
The dynamics, as described by the Hamiltonian constraint, is much more poorly understood%
---in particular, the full physical Hilbert space is not known---and we shall see that this may be 
important for our questions.

In loop quantum gravity the evidence is mixed, but there are at least some indications of 
lower-dimensional behavior at short distances.  This possibility was first noted by Modesto 
\cite{Modesto}, who pointed out that the average area in loop quantum gravity could be 
written in the form
\begin{align}
\langle{\hat A}_{\ell}\rangle 
  = \frac{\sqrt{{\ell}^2({\ell}^2+{\ell}_p^2)}}{\sqrt{{\ell}_0^2({\ell}_0^2+{\ell}_p^2)}}
  \langle{\hat A}_{{\ell}_0}\rangle
\label{d2}
\end{align}
where $\ell$ is a (quantized) length and $\ell_p$ is the Planck length.  For large $\ell$ 
this is just the ordinary scaling of area, but for small $\ell$ it differs, suggesting a change
in scaling dimensions of section \ref{geomd}.  If we translate this behavior into a 
scale-dependent effective metric and, as in the asymptotic safety program, use
this metric to determine a heat kernel and a spectral dimension, we find a flow
from $d_S=4$ at large scales to $d_S=2$ at small scales.  This analysis has been 
extended to the case of spin foams, the ``paths'' in the loop quantum gravity version
of a path integral, and a similar pattern of dimensional reduction appears \cite{Modestox}.

On the other hand, it is also possible to directly define a heat kernel and both Hausdorff 
and spectral dimension on a discrete complex such as a spin network or a spin foam \cite{Thurc}.  
Here the results are more ambiguous.  It seems that pure spin network states do not exhibit 
dimensional flow \cite{Thura}, and while superpositions that exhibit dimensional reduction 
can be constructed, the short-distance dimension depends on details of the superposition  
\cite{Thurb}.  The conclusion for physics may thus depend on the dynamics---exactly
which superpositions are allowed by the Hamiltonian constraint---rather than merely on
the kinematics.  As a possible first step in this direction, dimensional reduction has been
investigated in loop quantum cosmology, a ``minisuperspace'' theory in which all but
a few of the degrees of freedom of loop quantum gravity are frozen out \cite{Ronco,Miel}.
Here modifications of the full constraint algebra lead to modified dispersion relations 
(see section \ref{moddisp}).  The results are still somewhat ambiguous, depending on 
model-dependent choices for particular quantum corrections, but the simple choices yield 
spectral and a thermodynamic dimensions $d_S = d_{th2} = 2.5$ \cite{Ronco} or $d_S=1$ 
\cite{Miel} at short distances.  

\subsection{The short distance Wheeler-DeWitt equation \label{WdW}}

Inspired by the Hamiltonian formulation of the theory  \cite{ADM}, early attempts to 
canonically quantize general relativity led to the Wheeler-DeWitt equation \cite{DeWittb},
\begin{align}
\left\{ 16\pi\ell_p^2G_{ijkl}\frac{\delta\ }{\delta q_{ij}} \frac{\delta\ }{\delta q_{kl}}
    - \frac{1}{16\pi\ell_p^2}\sqrt{q}\,{}^{(3)}\!R\right\}\Psi[q] = 0
\label{d3}
\end{align}
where $q_{ij}$ is the spatial metric at a fixed time and $G_{ijkl} =\frac{1}{2}q^{-1/2}%
\left( q_{ik}q_{jl} + q_{il}q_{jk} - q_{ij}q_{kl}\right)$ is the DeWitt  supermetric.  This is 
almost certainly not quite the right way to quantize gravity---we don't know how to 
really make sense of the operators or construct a properly gauge-fixed inner 
product---but it is generally believed that the eventual quantum theory will contain 
something close to the Wheeler-DeWitt equation, at least as an approximation.  So 
if quantum gravity leads to short distance dimensional reduction, that fact ought to be 
evident here as well.

The wave function $\Psi[q]$ contains information about the metric at all scales.  To
focus in on small scales, we can look at the strong coupling limit $\ell_p\rightarrow\infty$.
As Isham first pointed out \cite{Isham}, this is also an ultralocal limit: the only term involving 
spatial derivatives of the metric drops out.\footnote{Strictly speaking, the theory is not quite 
ultralocal even in this limit, since the diffeomorphism constraint, which \emph{does} involve 
spatial derivatives, must still be imposed.}  This limit was studied extensively in the 1980s  
(see \cite{Carlipb} for references).   As the Planck length becomes large, particle horizons
shrink and light cones collapse to timelike lines, leading to the decoupling of neighboring 
points and the consequent ultralocal behavior \cite{Henneaux}.  In the completely 
decoupled limit, the solution at each point is essentially a Kasner space \cite{Helfer}, 
while for large but finite $\ell_p$ solutions exhibit BKL behavior \cite{BKL,BKLb,HUR},
looking locally Kasner but with chaotic ``bounces'' that change the Kasner axes and
exponents.

Now, Kasner space is certainly four-dimensional.  But as noted in section \ref{geomd},
geodesics at early times essentially see only one of the spatial dimensions, and the 
dimensional estimator $d_{geod}$ shrinks to $d_{geod}=2$ \cite{Carlipa,Carlipb}.  This
behavior has also been  noticed in a completely different astrophysical context
\cite{Chicone}.  As discussed in \cite{Carlipb}, it is plausible that the spectral dimension
exhibits a similar behavior.  The heat kernel for Kasner space is of the form
\cite{Futamase,Berkin}
\begin{align}
K(x,x;s) \sim \frac{1}{4\pi s^2}\left[ 1 + \frac{a}{t^2}\,s + \dots \right] .
\label{d4}
\end{align}
For a fixed time $t$, one can always find $s$ small enough that the first term dominates,
giving a ``microscopic'' spectral dimension $d_S=4$.   But for a fixed return time $s$, 
that is, a fixed scale at which one is measuring the dimension, there is always a time $t$ 
small enough that the second term dominates, giving an effective spectral dimension of 
$d_S=2$.

The key feature in this analysis is ``asymptotic silence'' \cite{HUR}, the collapse of light 
cones and the corresponding decoupling of neighboring points.  I will return to this 
issue in section \ref{silence}, but for now it is worth noting that the same phenomenon 
appears in a variety of other approaches to quantum gravity, from loop quantum cosmology 
to noncommutative geometry.

\subsection{Modified dispersion relations and noncommutative geometry \label{moddisp}}

The thermodynamic properties of a gas of particles depend on the dispersion relations
$E=E(p)$.   There are many reasons one might try to alter the standard dispersion relations, 
most of them unrelated to gravity.  But some research directions in quantum gravity---the
application of noncommutative geometry, for instance, or generalizations of the uncertainty 
principle to allow for a minimum length---lead naturally to such modifications.

It has been known for many years that modified dispersion relations can alter thermodynamic 
behavior \cite{Lubo}.  Inspired by the string theory results of section \ref{str}, several 
investigations in the early 2000s briefly mentioned the possibility that this might lead to 
dimensional reduction \cite{Rama,KGb,Mag}.  A concentrated focus on dimensional flow came 
more recently, though, starting, I believe, with an analysis by Husain, Seahra, and Webster of 
``polymer quantization,'' a loop-quantum-gravity-inspired modification of the ordinary rules 
of quantum mechanics \cite{Husain}.  Here, the thermodynamic dimension $d_{th2}$ was 
shown to drop to $d_{th2}=5/2$ at high temperatures.  Not long after, thermodynamic 
behavior was investigated for a particular noncommutative spacetime, Snyder space 
\cite{Nozari}, where it was shown that at high temperatures, $d_{th2}=d_{th3}=d_{th4}=2$.  
Thermodynamic dimensions were analyzed more systematically in \cite{Amelino} for a 
variety of dispersion relations, and it was demonstrated that the various dimensions 
of section \ref{thermd} could all differ from each other, and could differ from the spectral 
dimension as well.

Given a set of modified dispersion relations, the spectral dimension is closely related to 
the thermodynamic dimensions, although they are not quite equivalent \cite{Amelino}.  
Roughly speaking, a dispersion relation $E^2 - f(p^2)=0$ translated into a d'Alembertian
$\partial_t{}^2 - f(\nabla^2)$, which can be Wick rotated and inserted into (\ref{a4}) to 
obtain a spectral dimension.  Conversely, given \emph{any} energy dependence of a 
spectral dimension, one can reconstruct a dispersion relation that reproduces that 
dependence \cite{Sotiriou}.  This gives us an uncomfortable amount of freedom---we 
can choose any scale dependence of dimension we like, and build by hand a suitable 
dispersion relation---and it compels us to focus on dispersion relations that have a strong 
independent rationale.  

One such rationale is the polymer quantization of \cite{Husain}, in which dimensional 
flow arises from the nonstandard inner product of loop quantum gravity (although 
not all approaches to polymer quantization lead to this effect \cite{AL}).  Another rationale 
comes from the possible role of simple noncommutative geometries in quantum gravity, 
as in \cite{Nozari}, and their implied dispersion relations.  For instance, a form of 
``doubly special relativity''---a deformation of special relativity that preserves frame 
independence while introducing an invariant energy scale---leads to a spectral dimension 
that falls to $d_S=2$ at high energies, but the result depends on a free parameter 
\cite{Gubitosi}.  A similar reduction of spectral dimension occurs another popular model of 
noncommutative geometry, $\kappa$-Minkowski space, but the limiting value again depends 
on a nonunique choice, now of a generalized Laplacian \cite{Benedettib,Arzano}.   It 
has recently been argued that the static potential between two charges becomes constant at 
very small distances in $\kappa$-Minkowski space, corresponding roughly to a Greens function 
dimension $d_G\le3$ \cite{AK}.  Another model with a group-valued momentum space, 
inspired by exact results from three-dimensional gravity, also gives $d_S=2$ at high 
energies \cite{Nettel}.

On the other hand, we can instead start with the more formal approach to noncommutative 
geometry based on the spectral action developed Connes and his collaborators \cite{Connes}.
Here it appears that the spectral dimension is $d_S=0$ \cite{Alkofer,Vassil}.   A variation of the 
spectral action exists, though, that leads to a spectral dimension that fall from $d_S=4$ at 
large scales to $d_S=2$ at small scales, the ``conventional'' behavior we have seen 
elsewhere \cite{Kurkov}.

\subsection{Minimum length}

A frequent theme in quantum gravity is the possible existence of a minimum length 
\cite{Garay,Hossenfelder}. Heuristically, if one tries to probe too small a length, one needs so
much energy that the region being probed collapses into a black hole and becomes unobservable.  
While a minimum length is not a universal feature of quantum gravity, it is a common one, so 
it is interesting to see whether this is enough to imply some sort of short distance dimensional 
reduction.  

Modesto and Nicolini \cite{ModNic} have argued that one effect of a minimum length should 
be to smear out the starting point of the diffusion process described by (\ref{a4}), replacing 
the initial condition $K(x,x';0) = \delta(x-x')$ with a Gaussian distribution
\begin{align}
K(x,x';0) = \left(\frac{1}{4\pi\ell^2}\right)^{d/2}e^{-|x-x'|^2/4\ell^2}
\label{d5}
\end{align}
where $\ell$ is the minimum length.  The resulting spectral dimension drops from $d_S=d$ at
large scales to $d_S=d/2$ at the minimum length scale.  Padmanabhan, Chakraborty, and
Kothawala \cite{PadChak} have looked at scaling dimensions for a geodesic ball with a 
``quantum metric'' that incorporates a minimum length, finding that for any large-scale 
dimension, the box-counting dimension of section \ref{geomd} falls to $d_b=2$ at small scales.
Starting from the flow of spectral dimension in causal dynamical triangulations, Coumbe has 
suggested a  similar scale-dependent rescaling of length as a mechanism for dimensional
reduction \cite{Coumbey,Coumbez}.  A minimum length may also be incorporated through a 
``generalized uncertainty principle'' \cite{Hossenfelder}, which alters the Heisenberg commutation 
relations.  Such a change leads to modified dispersion relations, as in section \ref{moddisp}, 
so, not surprisingly, dimensional reduction appears at short distances  \cite{Husain,Gubitosi}.   
Maziashvili has also argued that the finite resolution due to a minimum length causes the 
box-counting dimension $d_b$ to decrease at small distances \cite{Maz}.

\subsection{Modified gravity}

Yet another approach to quantum gravity is to modify the Einstein-Hilbert action to make 
the theory renormalizable.  One popular example is ``Ho{\v r}ava-Lifshitz gravity'' \cite{Horavaa}, 
a model that sacrifices Lorentz invariance in exchange for (possible) renormalizability.  Here,
a generalization of the spectral dimension flows from $d_S=4$ at low energies to $d_S=2$ at 
high energies \cite{Horavab}, as do the thermodynamic dimensions $d_{th1}$ and $d_{th2}$ 
of section \ref{thermd} \cite{Alencar}.  Similarly, in curvature-squared models \cite{Stelle}, 
which are renormalizable but very probably nonunitary, both the Green function dimension 
$d_G$ \cite{Gegenberg} and a generalized spectral dimension  \cite{Calcagnic} fall to $d=2$.  In 
certain nonlocal renormalizable models, the spectral dimension again falls at high energies, 
now by an amount that depends on particular choices of form factors \cite{Modestob}.

While these results may be taken as further evidence for short distance dimensional reduction, 
they may also serve as cautionary notes.  The models I have described are defined on 
ordinary four-dimensional manifolds, and while the reduction of spectral dimension takes
place at high energies, it is essentially classical.  The same might be said for some of the
models of modified dispersion relations of section \ref{moddisp}.   The moral, perhaps, is
once again that one must be very careful about exactly what one means by ``dimension''%  
---different choices of how to measure dimension can give different results, which may not 
always reflect our intuition of what the dimension ``really'' is.

\subsection{Spacetime foam \label{foam}}

Many years ago, Wheeler suggested that quantum fluctuations of the geometry and 
perhaps the topology of spacetime should lead to a ``foamy'' structure at the Planck scale
\cite{Wheeler}.  A complete understanding of such behavior would certainly require a full 
quantum theory of gravity, but in the spirit of ``stochastic gravity'' \cite{Hub} we might try to
approximate spacetime foam by stochastic classical fluctuations.  Several attempts in
this direction have hinted at small scale dimensional reduction.  

One of the earliest proposals, due to Crane and Smolin \cite{Crane,Craneb}, models
spacetime foam as a collection of virtual black holes.  If the distribution of such black holes
is scale invariant and sufficiently dense, the spacetime outside their horizons becomes
fractal, with a Hausdorff dimension and a scaling dimension for Greens functions of
the form $d_H=d_G = 4-\varepsilon$, where $\varepsilon$ depends on the (unknown)
details of the distribution.  A few years later, Haba looked at the behavior of scalar Greens
functions in a spacetime with a stochastically fluctuating metric with a scale-invariant
spectrum of fluctuations \cite{Haba,Habab}.  The result is again a reduction of $d_G$ by 
an amount that depends on the details of the distribution.  Both proposals were motivated
in part by the hope that metric fluctuations might tame the infinities of quantum field
theory, and both have at least a heuristic connection to the asymptotic safety program
of section \ref{asafe}.  The minimum length scenario of Padmanabhan et al.\
\cite{PadChak} is in some sense a variation of the Crane-Smolin model, since the
``quantum metric'' can be viewed as generating a foamy structure of ``holes'' in
spacetime. 

A rather different stochastic treatment of vacuum fluctuations appears in \cite{Carlipd}.
As we have recently come to understand \cite{Fewster,Fewsterb}, quantum fluctuations 
of the vacuum have a highly non-Gaussian distribution, with a ``fat tail'' of large positive 
energy fluctuations.  These positive fluctuations focus null geodesics, and can be amplified
by the nonlinearities of the Raychaudhuri equation.  The result is a phenomenon known
in statistics as  ``gambler's ruin'':  large positive fluctuations are rare, but once a large enough 
fluctuation occurs, subsequent negative fluctuations are too small to reverse its effect.  In two 
spacetime dimensions, it was shown in \cite{Carlipd} that these fluctuations collapse light 
cones at scales near the Planck length, arguably leading to a kind of universal short distance 
asymptotic silence of the sort discussed in section \ref{WdW}.  It seems likely that this 
result can be extended to higher dimensions; see \cite{Carlipe} for a first step.

\subsection{Multifractional geometry \label{multif}}

Multifractional geometry is not so much a model of quantum gravity that predicts dimensional
reduction, but rather a broad mathematical framework that can naturally incorporate theories
with scale-dependent dimensions.   The formalism, based on fractional calculus, has been
developed extensively by Calcagni and collaborators \cite{Calcagni,Calcagnix,Calcagniy}, and 
can provide a setting for writing down particular models of dimensional reduction that capture
key features of other quantization programs.  As we shall see in section \ref{test}, a class of
multifractional models currently offer the strongest observational constraints on dimensional
reduction.  The framework also allows for the construction of models in which dimensional 
flow is related to an inherent uncertainty in length, or ``spacetime fuzziness,'' encoded in
stochastic properties of the measure \cite{ACR}.

\section{Common threads? \label{threads}}

We have seen that many approaches to quantum gravity show indications of dimensional 
reduction near the Planck scale.  Taken individually, none of these hints is terribly 
convincing.  Perhaps the best evidence comes from asymptotic safety, in 
which the argument for two-dimensional behavior at the ultraviolet fixed point is compelling, 
and causal dynamical triangulations, in which the evidence for flow of the spectral 
dimension is extremely strong.  But for this evidence to be truly persuasive, we would have 
to actually know that quantum gravity has an interacting ultraviolet fixed point, as required by
asymptotic safety, or that causal dynamical triangulations has the right continuum limit;
that is, we would have to know how to quantize gravity.

Taken as a body, though, these hints become quite a bit more compelling.  It seems rather unlikely 
that so many different approaches to quantum gravity would converge on the same result 
merely by accident.  If this convergence is more than coincidence, though, it ought to be
possible to to find a common thread, a single origin for dimensional reduction that is 
shared by all of the various approaches.

This is not easy.  Even for the much more mature subject of black hole thermodynamics, we 
don't really understand why so many approaches to quantization lead to the same expressions
for temperature and entropy, although there is some evidence that this behavior can be traced 
back to a common symmetry \cite{Carlipf,Carlipg}.  Moreover, as we have seen, different 
hints of dimensional flow employ different definitions of dimension, which need not
be equivalent.  If we are willing to be speculative, though, there have been suggestions of 
two possible common threads that could explain the ``universality'' of dimensional reduction.

\subsection{Scale invariance}

The fundamental premise of asymptotic safety is that the renormalization group flow of
quantum gravity has a non-Gaussian (that is, interacting) ultraviolet fixed point.  Such
a fixed point is, by definition, a point at which the theory becomes scale invariant.  This,
in turn, implies dimensional reduction, essentially because it is only in 
two dimensions that Newton's constant is dimensionless.

Causal dynamical triangulations arguably exhibits a similar scale invariance.  There is
growing evidence that a continuum limit occurs at a second order phase transition
\cite{AGG,ACG}, and at such a transition any theory becomes scale invariant \cite{scale}.
This result may connect back to asymptotic safety: it has been argued that the continuum
limit may only exist if the renormalization group flow for causal dynamical triangulations itself
has an ultraviolet fixed point \cite{Coopermanb}.

Similarly, the two terms in the Wheeler-DeWitt equation (\ref{d3}) scale differently under
constant rescalings of the metric, but in the short distance limit only one of these remains,
leading to a global scale invariance.\footnote{This invariance applies only for 
\emph{constant} rescalings.  In quantum field theory, such scale invariance plus Lorentz 
invariance usually implies full conformal invariance \cite{scaleCFT}, but it is not 
clear whether this holds for quantum gravity.}  The Crane-Smolin ``foam'' model discussed in
section \ref{foam} can be implemented with a scale-invariant distribution of virtual black
holes, again giving scale invariance at short distances.  For causal sets and loop quantum
gravity, the situation is less clear, but Dittrich has proposed that loop quantum gravity
and similar background-independent formulations should obey a set of consistency 
conditions under coarse-graining that are similar to the requirement of an interacting 
ultraviolet fixed point in asymptotic safety \cite{Dittrich}.

Amelino-Camelia, Arzano, Gubitosi, and Magueijo have argued forcefully for the fundamentality 
of scale invariance \cite{AAGMB}.  The central feature of the observed fluctuations of the
cosmic microwave background (CMB) is its nearly scale-invariant spectrum.  This is usually
taken as evidence for inflation, but Amelino-Camelia et al.\ propose that it is instead a
sign that at short enough distances and high enough energies, the universe really is
scale invariant.  Such a scale invariance may also be useful elsewhere, addressing some 
mysteries in particle physics; see, for example, \cite{Morozov}. 

\subsection{Asymptotic silence \label{silence}}

The Planck length $\ell_p = \sqrt{\hbar G/c^3}$ depends on Planck's constant, Newton's
constant, and the speed of light.  The $\ell_p\rightarrow\infty$ limit of section \ref{WdW}
is often described as a strong coupling limit, but it can equally well be viewed as an
``anti-Newtonian'' $c\rightarrow0$ limit \cite{Henneaux}.  In this limit, light cones collapse
to lines, and nearby points become causally disconnected.  This kind of behavior was first 
noticed in cosmology, where it was given the name ``asymptotic silence'' \cite{silence,silenceb}.  
In a sense, this is a reduction to one dimension, but the process goes by way of two dimensions, 
since, as described in section \ref{WdW}, the decoupling of neighboring points leads to BKL 
behavior with $d_{geod}\sim 2$.

What makes this behavior intriguing is that it appears in many different contexts
in quantum gravity.  It has been suggested that short distance asymptotic silence arises
directly from quantum fluctuations of the vacuum stress-energy tensor \cite{Carlipd}, as
discussed in section \ref{foam}.  Eichhorn et al.\ have shown that causal sets that 
approximate continuum spacetimes exhibit something akin to asymptotic silence,  
with space breaking up at short distances into disconnected ``islands'' \cite{Eichhornb}.  
Sorkin (cited in \cite{Eichhornb}) has also observed that in an asymptotically silent spacetime, 
one might expect drastically reduced scattering at high energies, since nearby worldlines do
not meet.  This does, indeed, happen in causal set theory, but also in string theory 
scattering at fixed small angle \cite{Gross,Ooguri}.  The apparent reduction of the spectral
dimension to $d_S=0$ in the Connes spectral action, as described in section \ref{moddisp},
has been conjectured to be related to asymptotic silence, and a similar result occurs in
a more ``Lorentzian'' approach to formal noncommutative geometry as well \cite{Besnard,Bizi}. 
And in loop quantum cosmology, there is evidence that the quantum deformation of the
constraint algebra leads to asymptotic silence at high curvatures and high energies 
\cite{Mielb}.
 
\section{Can dimensional reduction be tested? \label{test}}

Ideally, we would resolve the question of dimensional flow by simply looking at experimental 
or observational evidence.  In practice, this is unfortunately very difficult: in most approaches 
that exhibit dimensional reduction, the  phenomenon takes place near the Planck scale, 
well outside the range of any ordinary experiment.  On the other hand, we \emph{do} have 
some results at the Planck scale from searches for other phenomena such as broken Lorentz 
invariance \cite{Mat,Kos,Liberati}, so the possibility of observational tests is not completely 
outrageous.

Indeed, broken Lorentz invariance is the first place we might look for the effect of a reduction 
in the number of spacetime dimensions.  Unfortunately, though, the strongest existing
limits come from searches for ``systematic'' symmetry-breaking \cite{Mat,Liberati}---that is, 
patterns coming from a single global preferred frame---while the effects of dimensional reduction 
are typically ``nonsystematic,'' varying rapidly in space and time.  In asymptotic silence, 
for instance, the preferred Kasner axis that picks out the local two-dimensional spacetime at 
any given point varies chaotically \cite{Chernoff,Kirillov}.  There are some observational 
limits on nonsystematic violations of Lorentz invariance---see \cite{Liberati} for a recent 
review---but these are much more limited, and have not yet been applied to the question 
of dimensional reduction.

But while we do not yet have observational evidence, several lines of investigation are currently
being pursued.
\begin{itemize}
\item{\bf Cosmology:}\\[1ex]
The expansion of the universe---especially if it includes an inflationary epoch---can stretch out
signals at the Planck length to sizes that might become observationally accessible.  Cosmology is
thus a natural place to look for signs of Planck-scale dimensional reduction.  In the absence of
a clear set of quantum gravitational predictions, we can use various simplified models to
describe dimensional reduction; ultimately, we will have to understand the extent to which
any predictions depend on the choice of model.

As noted in section \ref{moddisp}, a scale-dependent spectral dimension can be translated into
modified dispersion relations \cite{Sotiriou}.  Using this observation, Amelino-Camelia et al.\
show  in \cite{Magueijo} that a spectral dimension $d_S=2$ at short distances naturally
leads to a scale-invariant spectrum of fluctuations in the CMB even without inflation.  In
more restrictive contexts, related ideas appeared earlier in \cite{Magb,Mukoyama}, and 
similar results come from asymptotically safe cosmology \cite{Bonannox}. Typical
models of this sort give predictions for the running of the spectral index of CMB fluctuations
that may be testable \cite{AAGMB,Barrowb,Mielc}, and a start has been made on investigating 
other properties of various models \cite{Brig,RSb}.  It is not yet known, however, whether there 
are features that would uniquely distinguish models with dimensional reduction from more 
conventional cosmologies.  

In a different approach to modeling dimensional reduction, two dimensions of spacetime can 
be initially ``frozen'' and then joined by mode-matching to a fully dynamical four-dimensional 
spacetime \cite{Rinaldi}.  Here, observational signatures in the CMB seemed to be washed out 
by an ensuing period of inflation.   On the other hand, certain multifractional models predict 
log-periodic oscillations, and these are now strongly constrained by CMB measurements 
\cite{Calcagnid}.  

In a rather different context, there has been recent interest in the cosmological implications
of quantum tunneling from a two-dimensional spacetime to four dimensions.  In the language 
of Lorentz symmetry breaking, this is a ``systematic'' change, one with a single global choice
of preferred dimensions.  As such, it is not directly applicable to most of the dimensional 
flow scenarios discussed here, but the results might still have some relevance.  Tunneling 
events of this sort lead to anisotropic spacetimes, although the anisotropy can be 
diluted by subsequent inflation.  Signatures of such an anisotropy in the CMB have been 
investigated in \cite{Adamek,Graham,Blanco,Scargill}.

In addition to primordial quantum fluctuations of the CMB , we can look directly at the CMB 
spectrum itself.  This spectrum probes much larger length scales and lower energies---we are 
looking at the surface of last scattering rather than the primordial fluctuations---but it tests 
different dimensional estimators.  In one of the earliest application of the notion of thermodynamic 
dimension, Caruso and Oguri showed in 2009 that the dimension of space at this scale differs 
from $d_{th}=4$ by at most a part in $10^{-5}$ \cite{Caruso}.

\item{\bf Particle physics:}\\[1ex]
Short-distance dimensional reduction can have implications throughout particle physics.  The 
problem, as usual, is that the Planck scale is so remote from any scale we can  directly access.
To have any possibility of obtaining useful results, we must look at high energies, 
long time spans for effects to accumulate, and precise measurements.  Even with these tricks,
 we have not yet managed to probe Planck scale effects, but in some models we are getting 
 closer:
\begin{itemize}
\item The first applications of particle physics to determine dimension were designed merely to 
measure the Hausdorff dimension of spacetime at ``ordinary'' scales.  By considering the anomalous 
magnetic moment of the electron in a spacetime of fractal dimension, Svozil and Zeilinger 
concluded that our spacetime must have a dimension $d_H>4-5\times10^{-7}$  \cite{Zeilinger}.  
A similar calculation of the Lamb shift gives $d_H>4-3.6\times10^{-11}$ \cite{Schafer}.
\item One might hope that processes dominated by loops in Feynman diagram might be more
sensitive to dimensional reduction, since they involve virtual particles of arbitrarily high energies.
By looking at $B$-${\bar B}$ oscillations in spacetimes with scale-dependent Hausdorff
dimensions, Shevchenko has shown that observations are almost completely insensitive to values 
$2<d_H<5$ at energies of 300-400 GeV \cite{Shevchenko}, corresponding to distances on the 
order of $10^{-18}$\,m (still, of course, much larger than the Planck length).
\item The most extensive studies have looked at certain classes of multifractional models 
\cite{Calcagnie,Calcagnif}.  The strongest bounds come from modifications of the Lamb shift
and the anomalous magnetic moment of the electron, both of which are measured with great
precision.  Such observations constrain the characteristic length scale $\ell_*$ for dimensional 
reduction in these models: for a general class of models, Calcagni et al.\ find $\ell_*<10^{-17}$\,m, 
and for a specific choice of parameters, $\ell_*<10^{-27}$\,m.  (For a review, see \cite{Calcagni}.)
\item Tests of Lorentz invariance achieve remarkable sensitivity by allowing possible violations 
to accumulate over very long distances, for instance by looking at photons from distant galaxies.  
The asymptotic silence picture of section \ref{silence} suggests the same approach: the passage 
through successive random Kasner-like spaces might have an observable effect on photon 
propagation.  An investigation of this possibility is in progress.
\end{itemize}

\item{\bf TeV-scale gravity:}\\[1ex]
The basic reason that observational searches for dimensional flow are so hard, of course, is 
that the Planck energy so high.  One exotic possibility is that in the correct theory of gravity,
the true Planck energy might be much lower, perhaps near the TeV scale.  Such modifications
of general relativity were originally introduced to solve puzzles in high energy physics, and
typically require extra dimensions \cite{Arkanib,RS}, seemingly the opposite of our focus here. 
But the hints of dimensional  reduction in section \ref{ev} are largely independent of the number 
of starting dimensions, so we can begin with more than four dimensions and still ask about 
dimensional reduction  to $d=2$.  Alternatively, we might simply postulate that dimensional 
reduction takes place for unknown reasons at energies less than $E_p$.  Several models 
have been built to exhibit this kind of behavior \cite{Jourjine,Kaplunovsky,Jourjineb,Ashb,Hao,Dai}, 
and there are suggestions that such a scenario might also help with some problems in high 
energy physics \cite{Stoj,He}.

With this assumptions, tests become much easier, and some of the limits described above become
significant.  In fact, there is even a claim of observational evidence for dimensional reduction, in 
the form of unexpectedly high planarity in secondary particles from cosmic ray collisions in the 
atmosphere \cite{Anchor}.  Accelerator tests have been proposed \cite{He,Anchor,Litimb,Gerwick}, 
based on several different models of dimensional reduction, though so far no relevant deviations 
from standard physics have been found.  It has also been suggested that dimensional reduction 
at TeV scales could affect the outcome of potential observations of primordial gravitational waves 
\cite{MurStoj}, though the effects seem to be quite model-dependent \cite{SVW}.
\end{itemize}

\section{Implications for physics}

I will close with some speculation about the possible implications of short distance dimensional 
reduction for the future.  Let us assume for the moment that the phenomenon is real, and that
it takes place near the Planck scale.  What does this mean for physics?

Of course, much of the answer depends on the details.  The two possible common threads of
section \ref{threads}, scale invariance and asymptotic silence, suggest rather different
directions for physics.  Still, there are a few ``universal'' characteristics.

First of all, dimensional reduction would almost certainly have far-reaching implications for 
cosmology.  One can invent exceptions: a contracting universe that ``bounces'' well before 
reaching the Planck density, for instance, might never reach a scale at which dimensional reduction 
is important.  In most cosmological models, though, the very early universe would
have a dimension smaller than four.  This would be reflected not only in the geometry, but in the 
structure of the quantum vacuum,  the equations of state for matter, and thermodynamic relations.  
Whether these effects would now be observable would depend on the amount of inflation---enough 
inflation can wash out almost any microscopic signature---but certainly any cosmology that hopes 
to deal with the birth of the universe would have to incorporate the fact that the newborn universe   
started with a different number of dimensions.

Second, dimensional reduction would almost certainly affect our understanding of black holes
\cite{Grumiller,Mureika,Fallsb}.  In particular, black hole thermodynamics depends strongly
on dimension, and the dynamics of evaporation would certainly be affected by dimensional
reduction near the horizon.  This could have implications for astronomy, influencing estimates 
of the densities of primordial black holes.  It would certainly have implications for fundamental
issues like the information loss problem, which depend strongly on the dimension-dependent
behavior of black holes in their final stages of evaporation.

Third, it is conceivable that dimensional reduction could profoundly affect our understanding
of quantum field theory.  The infinities of QFT are dimension-dependent, and there have 
been suggestions that dimensional reduction to $d=2$ at high energies could eliminate 
ultraviolet divergences \cite{Haba,Habab,Stoj,Anchora}.  Like most work in this area, these 
ideas have not yet been developed very deeply, but they could prove important.

Finally, dimensional reduction would surely affect the way we look at quantum gravity.  At 
the very least, it would suggest new formulations in which our four (or more) macroscopic
dimensions are split into two sectors with different characteristic length scales.  If, in addition, 
the dimension at the smallest scales is $d=2$, as suggested by many of the approaches of 
section \ref{ev}, this might allow us to employ the powerful methods of two-dimensional 
conformal field theory \cite{CFT}.

As noted in \cite{Carliph}, a number of authors have looked at the effect of such a splitting 
of spacetime, although in the rather different setting of high energy (eikonal) scattering 
\cite{tHooft,Verlinde,Kabat}.  They begin with a metric that can be written locally as
\begin{align}
ds^2 = \ell_\parallel^2 g_{\mu\nu}dx^\mu dx^\nu +\ell_\perp^2 h_{ij}dx^idx^j
\label{e1}
\end{align}
with the added condition that ``transverse'' derivatives $\partial_i$ are small.  The
Einstein-Hilbert action is then approximately
\begin{align}
I \sim \frac{\ell_\perp^2}{\ell_p^2}\int d^2x d^2y \sqrt{h}
   \left( \sqrt{-g}R_g 
   + \frac{1}{4}\sqrt{-g}g^{\mu\nu}\partial_\mu h_{ij}\partial_\nu h_{kl}
    \epsilon^{ik}\epsilon^{jl}\right) ,
\label{e2}
\end{align}
an expression that looks very much like a two-dimensional action for the transverse metric 
$h_{ij}$.  The action (\ref{e2}) is not quite conformally invariant: the trace of its effective 
stress-energy tensor is $T \sim \Box \sqrt{h}$, while conformal invariance would require $T=0$.  
But the deviation can be small, for instance falling to zero at small $t$ in Kasner space.  
Using a trick introduced by Solodhukin \cite{Solodukhin}, we can extract an effective scalar 
field by setting $h_{ij} = (1+\varphi)\sigma_{ij}$ with $\det\sigma_{ij}=1$.  To lowest order, 
$\varphi$ then turns out to be a two-dimensional Liouville field with a central charge  
$c\sim A_\perp/\ell_p^2$.  This is still far from the whole story---in the language of
section \ref{test}, the metric (\ref{e1}) exhibits a ``systematic'' breaking of Lorentz
invariance, with a global choice of longitudinal and transverse directions---but it offers
an interesting place to start.  A more general ``2+2'' splitting can be found in \cite{Yoon},
though the dual length scales have not yet been incorporated.  It may also be possible to 
use the general multifractional techniques of Calcagni described in section \ref{multif}.

Or perhaps we should be more radical.  In \cite{AAGMB} it is proposed that we should
start with a fundamental conformally invariant two-dimensional theory, from which
the added dimensions of spacetime emerge after conformal symmetry-breaking.
This is so far only a dream (though it is oddly reminiscent of the Polyakov approach
to string theory \cite{Polyakov}), but it could be a dream worth pursuing.

\vspace{1.5ex}
\begin{flushleft}
\large\bf Acknowledgments
\end{flushleft}

This work was supported in part by Department of Energy grant
DE-FG02-91ER40674.

\end{document}